\documentclass{article}
 \usepackage{mathtools}
 \usepackage{ mathrsfs }

\usepackage{amssymb}
\usepackage{amscd}
\usepackage{bm}
\usepackage{graphics}
\usepackage{amsfonts}

\usepackage{amsthm}
\usepackage{amsmath}
\usepackage{latexsym}
\usepackage[all]{xy}
\usepackage{fancyhdr}
\usepackage{color}
\usepackage{cite}

\usepackage{latexsym,graphics,color}

\def\be{\begin{equation}}
\def\ee{\end{equation}}

\def\D{\mathcal D}
\def\dd{\mathbb D}
\def\ka{\kappa}

\def\B{\mathcal B}

\def\dd{\mathbb D}

\def\E{\mathcal E}

\def\M{\mathcal M}

\def\bc{\mathbb C}

 \def\tr{{\rm tr}}

\def\K{\mathcal K}

\def\L{\mathcal L}
\def\si{\sigma}
\def\js{\frac{1}{4}}
\def\ad{{\rm ad}}

\def\d{\partial}
\def\jp{\frac{1}{2}}

\def\ri{{\mathrm i}}

\def\al{\alpha}

\def\ri{{\mathrm{i}}}                   %
\def\1{{\mbox{\boldmath $1$}}}          %
\def\tr{\mathrm{tr\,}}                  %
\def\lm{\lambda}                        %

\def\jp{\frac{1}{2}}                    %
\def\al{\alpha}                         %

\definecolor{spec}{rgb}{0.0, 0.26, 0.15}

\def\Ad{{\rm Ad}}
\def\bm{\begin{pmatrix}}
\def\em{\end{pmatrix}}

\begin{document}

\begin{flushright}
{}~
  
\end{flushright}

\vspace{1cm}
\begin{center}
{\large \bf Strong integrability of the bi-YB-WZ model}

\vspace{1cm}

{\small
{\bf Ctirad Klim\v{c}\'{\i}k\footnote{E-mail: ctirad.klimcik@univ-amu.fr}}
\\
Aix Marseille Universit\'e, CNRS, Centrale Marseille\\ I2M, UMR 7373\\ 13453 Marseille, France}
\end{center}

\vspace{0.5 cm}

\centerline{\bf Abstract}
\vspace{0.5 cm}
\noindent   We identify the r-matrix governing the Poisson brackets of the matrix elements of the Lax operator of the bi-YB-WZ model. 
  
  \vspace{2pc}


\section{Introduction}

Strong integrability is the property of certain Lax integrable theories. It amounts to Poisson-commutativity of the integrals of motions which are obtained as the spectral invariants of the Lax operator. In this paper, we focus on the issue of the strong integrability of Lax integrable non-linear $\sigma$-models. 

\medskip

The epitome of the Lax integrable non-linear $\sigma$ model is the principal chiral model \cite{ZM}, which can be formulated on any quadratic Lie group. First examples of integrable deformations
of the principal chiral model were obtained in Refs. \cite{C81,Mad93,Fa96} for the group  $SU(2)$ and it took one more decade to understand the algebraic structure behind those theories which permitted to generalize them to live on every simple compact group target. In particular, the models
\cite{C81,Fa96} were generalized in this way in Refs. \cite{K02,K09,K14} in the framework of  so-called $\eta$-deformation  (or Yang-Baxter deformation) procedure, while the model \cite{Mad93} was generalized  in Ref.\cite{S14} in the framework of  $\lm$-deformations. Since then, many other developments have followed, proving  the Lax integrability  of   complex structure-induced deformations \cite{By}, of  homogeneous
Yang-Baxter deformations \cite{KMY11,BW,OvT}, of   deformations with
a WZ term \cite{DMV15,DHKM}, of coset spaces \cite{DMV13}, of  
combined $\eta$ and $\lm$ deformations \cite{SST15} or of  
coupled $\eta$ and $\lm$ deformations \cite{DLMV19a,DLMV19b,L19,BL19,GS17,GS18,GSS17}. Moreover, it was established in Refs.  \cite{HT,K15,SST15,K19}, that some of those integrable deformations are related by  Poisson-Lie T-duality \cite{KS95}. 

\medskip

The strong integrability of the principal model was proved in Ref. \cite{Ma86b}, that of the single and of the double Yang-Baxter deformations
\cite{K09,K14} was established in \cite{DMV13,DLMV16} and that
of the single Yang-Baxter deformation with  WZ term was proved in \cite{DMV15}. On the other hand, the strong integrability 
of the $\lm$-deformed $\sigma$-model \cite{S14} was demonstrated in Ref. \cite{HMS,ISST} and of the coupled $\lm$-deformed models
\cite{GS17,GS18,GSS17} in Ref. \cite{GSS19}. Finally, the construction of the Lax pairs and the proof of the strong integrability of the coupled $\eta$ and $\lm$ deformations \cite{DLMV19a,DLMV19b,L19} was completed in Ref. \cite{BL19}.

\medskip

Restricting our attention to the simple group targets and taking into account dynamical equivalences of the models induced by the Poisson-Lie T-duality, it appears that the double Yang-Baxter deformation with  WZ term \cite{DHKM,K19} is the only known Lax integrable $\sigma$-model for which the strong integrability was not yet established. The principal result of the present article consists in filling this gap and  proving the strong integrability of this particular $\sigma$-model.

\medskip

Recall that the double Yang-Baxter deformation with  WZ term, or, shortly, the bi-YB-WZ model, is a 3-parametric submodel
of the $3+r^2$
parametric  DHKM $\sigma$-model constructed in Ref. \cite{DHKM}.
The  DHKM model lives on the simple compact (connected and simply connected) group $K$ and the integer $r$ is the dimension of the Cartan torus of $K$.  In a particular case of the $SU(2)$ target, the DHKM deformation is therefore four-parametric and turns out to coincide with the so called Lukyanov model \cite{L12}. It was later shown in Ref. \cite{K19} that the choice of the values of the $r^2$ parameters has no impact on the first order dynamics of the DHKM model, because changing the values of them can be undone by a suitable canonical transformation\footnote{This statement is true up to some zero modes subtleties.}. Said in other words, the models with different sets of the $r^2$ parameters are T-dual to each other, the duality in question being the dressing coset generalization of the Poisson-Lie T-duality \cite{KS96b}. Thus,
from the point of view of the first order Hamiltonian dynamics, the bi-YB-WZ model characterized by the vanishing of all $r^2$ parameters is dynamically equivalent to the general DHKM model, whatever be the choice of the values of the $r^2$ parameters of the latter. 

\medskip

The second order Lagrangian of the general DHKM $\sigma$-model is quite a complicated object but, as it was shown in Ref. \cite{K19}, it can be rewritten in the following succinct manner in the bi-YB-WZ case corresponding to the vanishing of the $r^2$ parameters:
  {\small  \be S_{\rm bi-YB-WZ}(m)=\ka \int d\tau\oint \tr\biggl( m^{-1}\partial_+ m\frac{\al+e^{\rho_rR_{m}}e^{\rho_lR}}{\al-e^{\rho_rR_{m}}e^{\rho_lR}}m^{-1}\partial_- m\biggr) +\ka\int d^{-1}\oint \tr(m^{-1}dm,[m^{-1}\partial_\sigma m,m^{-1}dm]).\label{biYBWZ}\ee}
   We postpone detailed explanations of the notations used in Eq. \eqref{biYBWZ} to Section 2, for the moment we just stress that it is this succinct manner which opens the way for solving the problem of strong integrability, which was left open in Refs. \cite{DHKM,K19}.

  \medskip
  
 It is remarkable, that we are able to prove the strong integrability of the bi-YB-WZ model by a shortcut method,  avoiding  the formalism of auxiliary fields which were originally used  in Refs. \cite{DHKM,K19} to define the model.  In particular, we succeed to express the first order dynamics of the bi-YB-WZ model not in terms of the dressing coset as in Ref. \cite{K19} but in terms of a much simpler {\it non-degenerate} $\E$-model. Such simplification makes possible to prove the strong integrability quite effortlessly.

   \medskip
   
   The plan of the paper is as follows. In Section 2,
   we  provide an overview of the Yang-Baxter deformations constructed in Refs. \cite{K02,K09,K14,DMV15,DHKM}. In Section 3, we give the formula for the Lax connection in terms of the second order target space field $m$ (this is also a new result of the present work).  We review shortly  the theory of $\E$-models  in Section 4 and we interpret the first order dynamics of the bi-YB-WZ model in terms of a particular non-degenerate $\E$-model in Section 5.  In  Section 6, we explain what the strong integrability means in general and in Section 7 we prove the strong integrability of the bi-YB-WZ model by writing down explicitely the $r$-matrix governing the Poisson brackets of the matrix elements of the bi-YB-WZ Lax operator.
  In Section 8, we show that our general formula for the bi-YB-WZ $r$-matrix does yield the $r$-matrices already known in literature for some specific choices of the deformation parameters. Section 9 contains concluding remarks and an outlook.

   \section{ Yang-Baxter deformations}
   
As we have already said in Introduction, the bi-Yang-Baxter deformation of the WZNW model was constructed in Ref. \cite{DHKM} and the succinct form of its  action was obtained\footnote{Actually, the action of the bi-YB-WZ model obtained in \cite{K19} coincides with the expression \eqref{biYBWZbis} upon the field redefinition replacing the group valued field $m$ by its inverse.}  in Ref. \cite{K19}:
  {\small  \be S_{\rm bi-YB-WZ}(m)=\ka \int d\tau\oint \tr\biggl( m^{-1}\partial_+ m\frac{\al+e^{\rho_rR_{m}}e^{\rho_lR}}{\al-e^{\rho_rR_{m}}e^{\rho_lR}}m^{-1}\partial_- m\biggr) +\ka\int d^{-1}\oint \tr(m^{-1}dm,[m^{-1}\partial_\sigma m,m^{-1}dm]).\label{biYBWZbis}\ee}
  Here  $\al\in]-1,1[$ and  $\rho_l,\rho_r\in]-\pi,\pi[$ are deformation parameters and the real positive level $\ka$ is quantized as usual so that the WZ term exhibits the   $2\pi$ ambiguity.  Note that the case $\al=0$ corresponds to the standard WZNW model.
  
  \medskip
  
  The further  notations are as follows: $m(\tau,\sigma)$ is a $K$-valued field on the cylindrical worldsheet parametrized by a time $\tau$ and by an angular space variable $\sigma$, the chiral derivatives are defined as  $\d_\pm:=\d_\tau\pm\d_\si$, $\oint$ stands for the integration over $\sigma$ and the symbol tr denotes  the usual (negative definite) Killing-Cartan form.  The standard Yang-Baxter operator $R:\K\to\K$  is defined by \be RB^\al=C^\al, \quad RC^\al=-B^\al,\quad RT^\mu=0.\label{YBoperator}\ee
Here $T^\mu$ is a basis of the Cartan subalgebra of the Lie algebra $\K$ of $K$ and $B^\al,C^\al$ are given in terms of the  step generators of $\K^\bc$ as \be B^\al=\frac{\ri}{\sqrt{2}}(E^\al+E^{-\al}), \quad C^\al=\frac{1}{\sqrt{2}}(E^\al-E^{-\al}).\label{2}\ee   
Finally,
  the operator $R_m:\K\to\K$ is defined as
 \be R_m:=\Ad_m^{-1}R\Ad_m\label{Rm}.\ee 
  
  \medskip
  
All other Yang-Baxter deformations previously constructed in the literature are appropriate special limits of the bi-YB-WZ one \eqref{biYBWZbis}. In particular, this is the case for the
  integrable Yang-Baxter deformation of the WZNW model introduced in Ref. \cite{DMV15}. Several equivalent expressions were obtained for the action of this YB-WZ  deformation in Refs.\cite{K17},\cite{DDST18} and \cite{K19}. We reproduce here the   parametrization given  in \cite{K19}:
   { \be S_{\rm YB-WZ}(m)=\ka \int d\tau\oint \tr\biggl( m^{-1}\partial_+ m\frac{\al+e^{\rho_lR}}{\al-e^{\rho_lR}}m^{-1}\partial_- m\biggr) +\ka\int d^{-1}\oint \tr(m^{-1}dm,[m^{-1}\partial_\sigma m,m^{-1}dm]).\label{YBWZ}\ee}
   Note that this is the special case of the action \eqref{biYBWZbis} obtained by setting $\rho_r=0$.
   
   \medskip
   
 Furthermore, setting \be \rho_r=2\ka b_r,\quad \rho_l=2\ka b_l,\quad \al=e^{-2\ka a},\label{par}\ee
and taking limit $\ka\to 0$, we recover from  the action \eqref{biYBWZbis} the bi-Yang-Baxter integrable deformation of the principal chiral model introduced in \cite{K02,K14}:
 \be S_{\rm bi-YB}(m)=-\int d\tau\oint \tr\biggl( m^{-1}\partial_+ m \left(a+b_rR_m +b_lR\right)^{-1}m^{-1}\partial_- m\biggr).\label{biYB}\ee

 Taking moreover $b_r=0$, we recover the Yang-Baxter $\sigma$-model \cite{K02,K09}, which was historically the first constructed integrable Yang-Baxter deformation:
 \be S_{\rm YB}(m)=-\int d\tau\oint \tr\biggl( m^{-1}\partial_+ m \left(a  +b_lR\right)^{-1}m^{-1}\partial_- m\biggr).\label{YB}\ee

\section{Lax connection}

The Yang-Baxter operator $R$ satisfies two useful identities  
 \be \tr(x\ \!Ry)=-\tr(Rx\ \!y), \quad x,y\in\K,\label{Rantisymmetry}\ee
 \be [e^{\rho R}x,e^{\rho R}y]= e^{\rho R}\left( [e^{\rho R}x,y]+[x,e^{\rho R}y]-2\cos{(\rho)}[x,y]\right) +[x,y],\quad x,y\in \K, \quad -\pi<\rho<\pi.\label{id}\ee
Using Eqs. \eqref{Rantisymmetry} and \eqref{id},
  the equations of motions of the bi-YB-WZ model \eqref{biYBWZbis} can be expressed in the form
  \be \al\d_+Y_--\al^{-1}\d_-Y_+-\al[e^{\rho_r R}Y_+,Y_-]-\al^{-1}[Y_+,e^{\rho_r R}Y_-]+2\cos{(\rho_l)}[Y_+,Y_-]=0,\label{fieldequation}\ee
where 
  \be Y_\pm=\left(e^{\rho_r R}-\al^{\mp 1}
  e^{- \rho_l R_{m^{-1}}}\right)^{-1}\d_\pm mm^{-1},\label{Ydef}\ee
 Note that for deriving the equations of motion we have   also used the Polyakov-Wiegmann   formula \cite{PW}
\be I_{\rm WZ}(m_1m_2)=I_{\rm WZ}(m_1)+I_{\rm WZ}(m_2)-\int d\tau\oint  \tr(m_1^{-1}\d_+m_1 \d_-m_2m_2^{-1})+\int d\tau\oint  \tr(m_1^{-1}\d_-m_1 \d_+m_2m_2^{-1}),\label{PW}\ee
where 
the  Wess-Zumino term $I_{\rm WZ}(m)$ is conveniently written in terms of the "inverse" $d^{-1}$ of the de Rham operator as follows  
\be I_{\rm WZ}(m)= \int d^{-1}\oint \tr\left(m^{-1}dm\wedge[m^{-1}\partial_\sigma m, m^{-1}dm]\right).\label{WZterm}\ee

Set now 
\be L_\pm(\xi):=\left( e^{\rho_r R}\mp f_\pm(\xi)\right)Y_\pm.
 \label{Laxconnect}\ee
  Here the meromorphic functions $f_\pm(\xi)$ are defined as
  \be f_\pm(\xi)= \frac{1}{1\pm \xi}\left(\al^{\mp 1}e^{-\ri\rho_l}(\xi\mp 1)+4\frac{\cos{\rho_l}-\al^{\mp 1}\cos{\rho_r}}{\al-\al^{-1}}\right).\label{pdef}\ee
and they  satisfy the  identity  
 \be f_+(\xi)f_-(\xi)=f_+(\xi)f_-(-1)+f_+(+1)f_-(\xi)+1.\label{fid}\ee

\medskip

\noindent {\bf Theorem (Lax connection)}: {\it If the field $m(\tau,\sigma)$ solves the equations of motion \eqref{fieldequation} then the Lax connection \eqref{Laxconnect} is flat, that is, it holds  for every
  $\xi\in\bc$, $\xi\neq \pm 1$
     \be \partial_+L_-(\xi)- \partial_-L_+(\xi) -[L_+(\xi),L_-(\xi)]=0.\label{zerocurv}\ee}
     \begin{proof}
  With the help of the identities \eqref{id} and \eqref{fid}, we find easily
   \be \partial_+L_-(\xi)- \partial_-L_+(\xi) -[L_+(\xi),L_-(\xi)]=
 \left(e^{\rho_r R}+ f_-(\xi)\right)V_-
- \left(e^{\rho_r R}- f_+(\xi)\right)V_+,
 \label{223x}\ee   
 where 
 \be V_\pm =\d_\mp Y_\pm -[L_\mp(\mp 1),Y_\pm]\label{Vdef}\ee
Note that the equations of motion \eqref{fieldequation} can be written as
\be \al^{-1}V_+=\al V_-.\label{VV}\ee
    We have the following obvious identity (valid even off-shell)
    \be \d_+(\d_-mm^{-1})-\d_-(\d_+mm^{-1})-[\d_+mm^{-1},\d_-mm^{-1}]=0\label{224}\ee
    that can be rewritten by using Eqs.\eqref{Ydef} as
$$ \d_+\left((e^{\rho_rR }-\al e^{-\rho_lR_{m^{-1}}})Y_-\right)- \d_-\left((  e^{\rho_rR }-\alpha^{- 1}e^{-\rho_lR_{m^{-1}}})Y_+\right) $$ \be - \left[( e^{\rho_rR }-\al^{-1}e^{-\rho_lR_{m^{-1}}})Y_+,( e^{\rho_rR }-\al e^{-\rho_lR_{m^{-1}}})Y_-\right]=0.\label{226}\ee
    We find
    {\small $$ \d_\pm \left(e^{-\rho_lR_{m^{-1}}}Y_\mp\right)= e^{-\rho_lR_{m^{-1}}}\left(\d_\pm Y_\mp -[\d_\pm mm^{-1}, Y_\mp]\right)+\left[ \d_\pm mm^{-1},e^{-\rho_lR_{m^{-1}}}Y_\mp\right]=$$\be=e^{-\rho_lR_{m^{-1}}}\left(\d_\pm Y_\mp- [( e^{\rho_rR}-\alpha^{\mp 1}e^{-\rho_lR_{m^{-1}}})Y_\pm,Y_\mp]\right)+  \left[ ( e^{\rho_rR}-\alpha^{\mp 1}e^{-\rho_lR_{m^{-1}}})Y_\pm,e^{-\rho_lR_{m^{-1}}}Y_\mp\right]. \label{228} \ee}
   Using Eqs. \eqref{228} and  \eqref{id}, we can rewrite the identity \eqref{224} as follows
 {\small \be    0=  \d_+(\d_-mm^{-1})-\d_-(\d_+mm^{-1})-[\d_+mm^{-1},\d_-mm^{-1}]=e^{\rho_rR}(V_-- V_+)-e^{-\rho_lR_{m^{-1}}}(\al V_--\al^{-1}V_+).\label{230}\ee}
Because we supposed that the field $m$ fulfils the equations of motion $\al V_-=\al^{-1}V_+$, we infer from the identity \eqref{230} that it holds
\be V_\pm=0.\ee
Inserting this information into Eq.\eqref{223x}, we conclude that the Lax connection is flat on-shell for every $\xi\neq\pm 1$, that is, it holds 
\be \partial_+L_-(\xi)- \partial_-L_+(\xi) -[L_+(\xi),L_-(\xi)]=0.\label{245}\ee 
\end{proof}
     In this way, we have proved the Lax integrability of the bi-YB-WZ model. To prove the strong (or Hamiltonian) integrability we have first to expose in the next section some useful  material about the $\E$-models.

     \section{Overview of $\E$-models}
     

The non-degenerate $\E$-models  were introduced in \cite{KS96a} as first order dynamical systems  which encapsulate the  Hamiltonian dynamics of certain T-dualizable non-linear $\sigma$-models. The $\E$-model is associated to the following
 data: 

\medskip

1) A Drinfeld double $D$, which is an even-dimensional Lie group equipped with a bi-invariant Lorentzian metric of the split signature $(d,d)$.  This metric naturally induces a non-degenerate symmetric ad-invariant  bilinear form $(.,.)_\D$ on the Lie algebra $\D$ of $D$.

\smallskip

2) An eponymous  linear operator $\E$ acting on the Lie algebra $\D$ of the double $D$ which has three important properties : i) it squares to the identity operator on $\D$, i.e. $\E^2=$ Id; ii) it is self-adjoint with respect to the bilinear form $(.,.)_\D$, i.e. $(\E x,y)_\D=(x,\E y)_\D$, $x,y\in \D$; iii) a symmetric  bilinear form on $\D$ defined as $(.,\E .)_\D$ is strictly positive definite. 

\medskip

Actually, the datum 2) can be reformulated equivalently as

\medskip

2') A half-dimensional subspace $\E_+\subset \D$ such that the restriction of the bilinear form $(.,.)_\D$ on $\E_+$ is strictly positive definite. 

\medskip

\noindent {\bf Remark 1:} {\small The subspace $\E_+\subset \D$ appearing in the datum 2') is the $(+1)$-eigenvalue eigenspace of the operator $\E$ appearing in 2). It then turns out that $\E_-\equiv \E_+^\perp$ is the $(-1)$-eigenvalue eigenspace of $\E$, the restriction of $(.,.)_\D$ on $\E_-$ is strictly negative definite and $\D$ can be written as the direct product $\D=\E_+\oplus \E_-$.}

\medskip

The phase space of the $\E$-model is the loop group $LD$ of the Drinfeld double. The symplectic form $\omega_{LD}$ and the Hamiltonian $H_\E$ are given by the expressions
\be \omega_{LD}:=-\jp \oint  (l^{-1}dl,\partial_\sigma(l^{-1}dl))_\D, \label{356}\ee
\be H_\E=\jp\oint  (\partial_\sigma ll^{-1},\E \partial_\sigma ll^{-1})_\D\label{359}\ee
and the corresponding first-order action of the $\E$-model reads \cite{KS96a}
\be S_\E(l)=\jp\int d\tau\oint (\partial_\tau ll^{-1},\partial_\sigma ll^{-1})_\D+\frac{1}{4}\int d^{-1}\oint (dll^{-1}\stackrel{\wedge}{,}[\partial_\sigma ll^{-1}, dll^{-1}])_\D-\jp\int d\tau\oint (\partial_\sigma ll^{-1},\E\partial_\sigma ll^{-1})_\D.\label{360} \ee
 The equations of motions derived from the action  \eqref{360} can be written either as
 \be \partial_\tau ll^{-1}=\E\partial_\sigma ll^{-1}\label{216}\ee
 or as
 \be \partial_\tau j=\partial_\sigma(\E j)+[\E j,j],\label{218}\ee
 where
 \be j(\sigma):= \partial_\sigma l(\sigma)l(\sigma)^{-1}.\ee
 The Poisson brackets of the components of the $\D$-valued current $j(\sigma)$ can be derived from the
 symplectic form $\omega_{LD}$ and they read
 \be \{(j(\si),z)_\D,(j(\si'),z')_\D\}=(j(\si),[z,z'])_\D+(z,z')_\D
 \partial_\sigma\delta(\sigma-\sigma'), \quad z,z'\in\D.\label{223}\ee
 
 \medskip

 Let $B\subset D$ be a half-dimensional  isotropic  subgroup of the Drinfeld double; the isotropy  means that the restriction of the form $(.,.)_\D$ on the Lie subalgebra $\B\subset\D$ vanishes identically. As it was shown in \cite{KS97, K15}, the $\E$-model action 
 \eqref{360} describes the  first order Hamiltonian dynamics
 of the non-linear $\sigma$-model living on the space of right cosets $D/B$. The second order action of this $\sigma$-model is obtained from Eq.\eqref{360} by writing the $D$-valued field $l$ as 
 \be l=mb,\quad b\in B,\label{379}\ee 
 and subsequently by solving the $B$-valued field $b$ out. The result is given by the formula
$$ S_\E(m)=+\frac{1}{4}\int d^{-1}\oint \biggl(m^{-1}dm,[m^{-1}\partial_\sigma m,m^{-1}dm]\biggr)_\D+$$\be +\frac{1}{4} \int d\tau\oint \biggl( m^{-1}\partial_+ m,P_m(-\E) m^{-1}\partial_- m\biggr)_\D-\frac{1}{4} \int d\tau\oint \biggl(P_m(+\E)m^{-1}\partial_+ m,m^{-1}\partial_- m\biggr)_\D , \label{219}\ee
 where the $D/B$-valued $\sigma$-model field is parametrized by (possibly, the collections of local) sections $m\in D$ of the bundle 
  $D\to D/B$.  
The projectors $P_m(\pm\E)$ are defined by their common  image $\B$ and by their respective kernels  Ad$_{m^{-1}}\E_\pm=$ (Id$\pm$Ad$_{m^{-1}}\E$Ad$_{m})\D$. 

The equations of motion corresponding to the action  \eqref{219} have the form of the zero curvature condition in the Lie algebra $\B$
\be \d_+ K_- -\d_- K_+-[K_+,K_-]=0, \label{385}\ee
where the $\B$-valued currents $K_\pm$ are given by
\be 
K_\pm=-P_m(\pm\E)m^{-1}\d_\pm m.\label{388}\ee
For deriving  the equations of motions, the following identity is particularly useful
\be P_m(\pm\E)^\dagger=1-P_m(\mp \E),\label{390}\ee
where the symbol $\dagger$ denotes taking the adjoint operator with respect to the bilinear form $(.,.)_\D$.

\medskip

Note also, that if $l=mb$ is a solution of the first order equations of motion \eqref{216} and $m$ the corresponding solution of Eqs. h\eqref{385}, \eqref{388} then it holds
\be \d_\pm bb^{-1}=K_\pm=-P_m(\pm\E)m^{-1}\d_\pm m.\label{397}\ee
This relation will be needed in Section 7.
\bigskip

  \section{Bi-YB-WZ model as $\E$-model}

  The bi-YB-WZ model \eqref{biYBWZbis} was constructed in Ref. \cite{K19} as a particular {\it degenerate} $\E$-model (i.e. the dressing coset) based
  on the Drinfeld double $\dd=K^\bc\times K^\bc$. In the present section, we show that the same  bi-YB-WZ model can be constructed in a much simpler way as the standard {\it non-degenerate} $\E$-model based on a smaller Drinfeld double $D=K^\bc$.  Let us  describe this new simpler construction in detail. 

 To recover the bi-YB-WZ $\sigma$-model \eqref{biYBWZbis} as the special case of the formula \eqref{219}, we have to consider the following particular $\E$-model data:
 
 \medskip
 
 1) The Drinfeld double $D$  is the complexified group $K^\bc$ and the bilinear form $(.,.)_\D$ is given by the formula
  \be \left(z,z'\right)_\D:=\frac{4\ka}{\sin{(\rho_l)}}  \Im\tr\left(e^{\ri \rho_l}zz'\right),\quad z,z'\in \K^\bc.\label{498a}\ee
Here the symbol  $\Im$ means  the imaginary  part of a complex number and  $\ka,\rho_l$ are the parameters appearing in the
action \eqref{biYBWZbis}.  

\smallskip

2') The subspaces $\E_\pm$ are given by
\be   \E_\pm=\left\{(\al^{\pm 1}-e^{-\ri\rho_l}e^{-\rho_rR})x, \  x\in\K\right\}.\ee
Here $R$ is the Yang-Baxter operator and the real parameters $\al$ and $\rho_r$ are again those appearing in the action \eqref{biYBWZ}. 

\bigskip

The half-dimensional isotropic subgroup $B$ is obtained by exponentiation of the Lie subalgebra $\B\subset K^\bc$ defined as
\be \B=\left\{ \frac{e^{-\ri \rho_l}-e^{-\rho_l R}}{\sin{\rho_l}}y, \ y\in\K\right\}.\label{387}  \ee
The fact that the subspace defined by \eqref{387} is the Lie subalgebra of $\K^\bc$ is the consequence of the properties of the Yang-Baxter operator, namely of the identity \eqref{id} rewritten as
 \be \left[\frac{e^{-\ri \rho_l}-e^{-\rho_l R}}{\sin{\rho_l}}x,
\frac{e^{-\ri \rho_l}-e^{-\rho_l R}}{\sin{\rho_l}}y\right]=
\frac{e^{-\ri \rho_l}-e^{-\rho_l R}}{\sin{\rho_l}}[x,y]_{R,\rho_l},\quad x,y\in \K.\label{Bbracket}\ee
Here $[.,.]_{R,\rho_L}$ is an alternative Lie bracket on the vector space $\K$  defined in terms of the standard Lie bracket $[.,.]$ and of the Yang-Baxter operator as
\be [x,y]_{R,\rho_l}:=\left[\frac{\cos{ \rho_l}-e^{-\rho_l R}}{\sin{\rho_l}}x,y\right]+\left[x,\frac{\cos{ \rho_l}-e^{-\rho_l R}}{\sin{\rho_l}}y\right].\label{Rrhobracket}\ee

\noindent {\bf Remark 2:} {\small The group $B$ turns out to be the semi-direct product of a suitable real form of the complex Cartan torus $\mathbb T^\bc$ with the nilpotent subgroup $N\subset K^\bc$ generated  by the positive step operators
$E^\al$.  It is also worth noting that in the limit $\rho_l\to 0$ the alternative commutator \eqref{Rrhobracket}
becomes 
\be [x,y]_{R,\rho_l\to 0}=[x,y]_R:=[Rx,y]+[x,Ry]. \label{Rbracket}\ee
and the identity \eqref{Bbracket} becomes the standard Yang-Baxter identity
\be [(R-\ri)x,(R-\ri)y]=(R-\ri)[x,y]_R.\label{YBid}\ee}

\medskip
 
 The space $D/B$ turns out to be just the group $K$,
 therefore the field $m$ appearing in the second order action \eqref{219} is simply $K$-valued. We then find
 \be P_m(\pm\E)m^{-1}\d_\pm m=\left(e^{-\ri \rho_l}-e^{-\rho_l R}\right)\left(\alpha^{\pm 1} e^{\rho_rR_{m}}-e^{-\rho_lR}\right)^{-1}m^{-1}\d_\pm m\label{408}\ee
 and, taking into account also the property \eqref{Rantisymmetry}, we find tb             hat the action \eqref{219} becomes
    \be S_\E(m)=\ka \int d\tau\oint \tr\biggl( m^{-1}\partial_+ m\frac{\al+e^{\rho_rR_{m}}e^{\rho_lR}}{\al-e^{\rho_rR_{m}}e^{\rho_lR}}m^{-1}\partial_- m\biggr) +\ka\int d^{-1}\oint \tr(m^{-1}dm,[m^{-1}\partial_\sigma m,m^{-1}dm]).\label{biYBWZtris}\ee
  We observe that the action \eqref{biYBWZtris} coincides with the bi-YB-WZ action \eqref{biYBWZbis}. 
    
 \section{Strong integrability - general story}
 Before settling the problem of the strong integrability of the bi-YB-WZ model, we recall what the strong integrability means in general.
 
 \medskip
 
 A dynamical system is said to be Lax integrable if it exists a  {\it Lax pair} $(\L,\M)$ consisting of two operator-valued  functions on the phase space such that  the complete set of the first-order equations of motions of the system can be expressed in the Lax way
   \be \frac{d}{dt}\L=[\L,\M]\label{Laxpair}.\ee
   Here $[.,.]$ means the commutator of linear operators acting on some auxiliary vector space $V$.
   
   \medskip
   
  If the Lax condition \eqref{Laxpair} holds, it is evident that spectral invariants of the Lax operator $\L$ (typically traces of the powers of $\L$) are conserved quantities.
  
  \medskip
  
  If the system has many degrees of freedom, the auxiliary space $V$ must have big dimension in order that the complete set of equations of motion be expressed as in \eqref{Laxpair}. However, it exists a variant of the weak Lax integrability in which 
  the operators $\L,\M$ depend not only on the phase space variables but they are also  meromorphic functions of some auxiliary complex variable $\xi$ called  {\it spectral parameter}. 
  In this case the Lax condition with spectral parameter reads
     \be \frac{d}{dt}\L(\xi)=[\L(\xi), \M(\xi)]\label{Laxpairspectral}.\ee
     It is understood that the complete set of the first order equations of motion is obtained making
     to hold the relation \eqref{Laxpairspectral} for every non-singular value of the spectral parameter
     $\xi$. This in many cases permits to consider the auxiliary vector spaces $V$ of small dimensions.
  
  \medskip
  
   We know already that the spectral invariants of the Lax operator are conserved quantities but this fact does not mean automatically that Poisson bracket of every two spectral invariants vanishes. However, if this happens, the Lax integrability of the system is referred to as being {\it strong}.
   
   \medskip
   
   It is well-known (see e.g. \cite{Ma85,Ma86}), that a sufficient condition for the strong  Lax integrability is the existence of the so called $r$-matrix  which is an operator acting on the tensor product $V\times V$. Furthermore, the $r$-matrix  is meromorphic in two complex variables $\xi,\zeta$ and it may (though it need not) depend on the phase space variables\footnote{If it does depend on the phase space variables, it is called a {\it dynamical} $r$-matrix.}. This $r$-matrix must fulfil the following crucial relation
  \be \{\L(\xi)\otimes {\rm Id},{\rm Id}\otimes \L(\zeta)\}=[r(\xi,\zeta),\L(\xi)\otimes {\rm Id}]-[r^p(\zeta,\xi),{\rm Id}\otimes \L(\zeta)],\label{rLax}\ee
  where $\{.,.\}$ stands for the Poisson bracket and $r^p$ is the permuted $r$-matrix. More precisely,
   if $r$ can be written as
  \be r=\sum_\alpha A_\alpha\otimes B_\alpha \ee
  for some family of linear operators $A_\alpha,B_\alpha$ acting on $V$, then the notation $r^p$ means  
  \be r^p=\sum_\alpha B_\alpha\otimes A_\alpha.\ee

  \medskip 
 \section{Strong integrability - the bi-YB-WZ case}
 The fact that the bi-YB-WZ $\sigma$-model \eqref{biYBWZbis} has the first order Hamiltonian formulation in terms of the $\E$-model is particularly useful for establishing its strong integrability. To show this, we start by recalling the explicit formula for the Lax connection obtained in Section 3
 \be L_\pm(\xi):=\left( e^{\rho_r R}\mp f_\pm(\xi)\right)Y_\pm,
 \label{Laxconnectiontris}\ee
  \be Y_\pm=\left(e^{\rho_r R}-\al^{\mp 1}
  e^{- \rho_l R_{m^{-1}}}\right)^{-1}\d_\pm mm^{-1},\label{Udefbis}\ee
  \be f_\pm(\xi)= \frac{1}{1\pm \xi}\left(\al^{\mp 1}e^{-\ri\rho_l}(\xi\mp 1)+4\frac{\cos{\rho_l}-\al^{\mp 1}\cos{\rho_r}}{\al-\al^{-1}}\right).\label{pdefbis}\ee
  Recall that the  meromorphic functions $f_\pm(\xi)$ satisfy the useful identity \eqref{fid}
 \be f_+(\xi)f_-(\xi)=f_+(\xi)f_-(-1)+f_+(+1)f_-(\xi)+1.\label{pidbis}\ee
          
  The Lax pair operators $\L$ and $\M$ act on the auxiliary (loop) space
  $V=L\K$ as
  \be \L(\xi)\chi=\d_\si \chi+\jp[L_-(\xi)-L_+(\xi),\chi],\quad \M(\xi)\chi=-\jp[L_-(\xi)+L_+(\xi),\chi], \quad \chi(\si)\in L\K,\label{514}\ee
or, shortly,
         \be \L(\xi)=\d_\si+\jp\ad_{L_-(\xi)-L_+(\xi)}, \quad \M(\xi)=-\jp\ad_{L_-(\xi)+L_+(\xi)}.\label{504}\ee
         Indeed, 
         the
         zero curvature condition \eqref{zerocurv}
         then amounts just to the Lax condition
         \be \d_\tau\L(\xi)=[\L(\xi),\M(\xi)]\label{Laxcondition}\ee
         and we already know that in the bi-YB-WZ case it encodes the complete set of equations of motion of the model.

In order to establish the strong integrability of the bi-YB-WZ model, we  must calculate the matrix Poisson bracket of the Lax operator $\L(\xi)$ with itself as in the left-hand-side of Eq.\eqref{rLax}. This in turn boils down to the calculation of the Poisson brackets of the $\K$-valued currents $Y_\pm$.
The utility of the $\E$-model approach constitutes in the fact, that this Poisson $Y$-current algebra can be easily found from the general $\E$-model formula \eqref{223}
 because it holds
    \be \tr(Y_\pm(\sigma)x)= \left(j(\sigma),\frac{\al^{\pm 1}e^{\rho_rR}-e^{-\ri\rho_l}}{2\ka(\al-\al^{-1})}x\right)_\D,\quad x\in\K.\label{Ypm}\ee
  
To show the validity of the formula \eqref{Ypm}, we use the formulas \eqref{216}, \eqref{379}, \eqref{397} and \eqref{408}
$$(\E\pm 1)j=\d_\pm ll^{-1} =\d_\pm mm^{-1}+m\d_\pm bb^{-1}m^{-1}=$$\be=
 \d_\pm mm^{-1}-m\left(\left(e^{-\ri \rho_l}-e^{-\rho_l R}\right)\left(\alpha^{\pm 1} e^{\rho_rR_{m}}-e^{-\rho_lR}\right)^{-1}m^{-1}\d_\pm m\right)m^{-1}= 
 \left(e^{\rho_r R}-\al^{\mp 1}e^{-\ri \rho_l} \right)Y_\pm.\label{464}\ee
 We then find easily
 $$\left(j(\sigma),\frac{\al^{\pm 1}e^{\rho_rR}-e^{-\ri\rho_l}}{2\ka(\al-\al^{-1})}x\right)_\D=\frac{1}{4\ka }(j,(\E\pm 1)e^{\rho_rR}x)_\D  =\frac{1}{4\ka }((\E\pm 1)j,e^{\rho_rR}x)_\D=$$\be =\frac{1}{\sin{\rho_l}} \Im\tr\left(e^{\ri \rho_l}\left(e^{\rho_r R}-\al^{\mp 1}e^{-\ri \rho_l} \right)Y_\pm e^{\rho_rR}x\right)=\tr(Y_\pm x).\label{467}\ee
  Note that we have thus shown  
 \be j=\jp\left(  e^{\rho_r R}-\al^{-1}e^{-\ri \rho_l} \right)Y_+-\jp\left(  e^{\rho_r R}-\al e^{-\ri \rho_l} \right)Y_-.\label{468}\ee
 
   It is convenient to rewrite the Lax operator $\L(\xi)$ as
       \be \L(\xi)=\d_\si-\ad_{L(\xi)}  =\d_\si-\ad_{f(\xi)Y_1}-\frac{1}{4\ka}\ad_{(e^{\rho_r R}+h(\xi))Y}\label{Laxfh}\ee
         where
         \be Y_1:=2\ka(\al^{-1}Y_+-\al Y_-),\quad Y=2\ka(Y_+-Y_-),\label{Y1Y}\ee
         \be  \quad f(\xi):=\frac{f_+(\xi)+f_-(\xi) }{4\ka(\al-\al^{-1})},\quad  h(\xi):=-\frac{\al f_+(\xi)+\al^{-1} f_-(\xi)}{\al-\al^{-1}}.\label{530}\ee
       %
The reason for that is the fact  that the Poisson brackets involving the current components $Y$ and $Y_1$ have particularly simple form. Indeed, from Eqs.\eqref{Ypm} and \eqref{Y1Y} we find
\be \tr(Y_1 x)= (j,e^{-\ri \rho_l}x)_\D,\quad \tr(Yx)=  (j,e^{\rho_rR}x)_\D\label{477}.\ee
The  identities \eqref{223} and \eqref{id} then permit to calculate the   Poisson brackets of the $Y$-current algebra
  \be \{\tr(Y_1(\si_1) x),\tr(Y_1(\si_2) y)\}=  \ \!\tr\left(\left(2\cos{(\rho_l)}Y_1(\si_1)  -  e^{+\rho_rR}Y(\si_1)\right)[x,y]\right)\delta(\sigma_1-\sigma_2)- 4\ka\tr(xy)\delta'(\sigma_1-\sigma_2),\label{488}\ee
    \be \{\tr(Y(\si_1) x),\tr(Y(\si_2) y)\}=  \ \!\tr\left( Y(\si_1)\left( e^{-\rho_rR}[e^{\rho_rR}x,e^{\rho_rR}y]\right)\right) \delta(\sigma_1-\sigma_2)+ 4\ka\tr(xy)\delta'(\sigma_1-\sigma_2),\label{489}\ee
      \be \{\tr(Y_1(\si_1) x),\tr(Y(\si_2) y)\}=   \ \!\tr\left( Y_1(\si_1)[x,e^{\rho_rR}y]\right) \delta(\sigma_1-\sigma_2).\label{490}\ee
It is now straightforward to calculate the Poisson bracket of the matrix elements of the Lax operator 
{\small $$ \{\tr(\L(\xi)(\si_1)\ad_x),\tr(\L(\zeta)(\si_2)\ad_y)\} =$$$$=\left\{f(\xi)\tr(Y_1(\si_1)x)+\frac{1}{4\ka}\tr\left(Y(\si_1)(h(\xi)+e^{-\rho_rR})x\right),f(\zeta)\tr(Y_1(\si_2)y)+\frac{1}{4\ka}\tr\left(Y(\si_2)(h(\zeta)+e^{-\rho_rR})y\right)\right\}=$$$$=  f(\xi)f(\zeta) \ \!\tr\left(\left(2\cos{(\rho_l)}Y_1(\si_1)  -  e^{+\rho_rR}Y(\si_1)\right)[x,y]\right)\delta(\sigma_1-\sigma_2)+$$$$+\frac{1}{(4\ka)^2 }\tr\left(Y(\si_1)\left(e^{-\rho_rR}\left[\left(h(\xi)e^{\rho_rR}+1\right)x,\left(h(\zeta)e^{\rho_rR}+1\right)y\right]\right)\right)\delta(\sigma_1-\sigma_2)+$$
$$+\frac{1}{4\ka}\tr\left(Y_1(\si_1)\left(f(\xi)\left[x,\left(h(\zeta)e^{\rho_rR}+1\right)y\right]+f(\zeta)\left[\left(h(\xi)e^{\rho_rR}+1\right)x,y\right]\right)\right)\delta(\sigma_1-\sigma_2)+$$
\be+\frac{1}{4\ka}\left( \left(h(\xi)h(\zeta)+1-(4\ka)^2 f(\xi)f(\zeta)\right) \tr(xy)+h(\xi)\tr(ye^{\rho_rR}x)+h(\zeta)\tr(xe^{\rho_rR}y)\right)\delta'(\sigma_1-\sigma_2).\label{LxLy}\ee}

Now we look for the $r$-matrix $r(\xi,\zeta):L\K \otimes L\K\to L\K\otimes L\K$ to be inserted into the right-hand-side of the strong integrability condition \eqref{rLax}  to match the formula \eqref{LxLy}. We choose the following ansatz for the matrix elements of the  $r$-matrix $r(\xi,\zeta)$
\be  (\tr\otimes\tr)\Bigl((x\otimes y)r(\xi,\zeta)\bigl(\chi_1(\si_1)\otimes\chi_2(\si_2)\bigr)\Bigr)= \tr\Bigl([x,\chi_1(\si_1)]\hat r(\xi,\zeta)[y,\chi_2(\si_2)]\Bigr)\delta(\sigma_1-\sigma_2),\  x,y\in\K, \   \chi_{1,2}\in L\K,\label{ransatz}\ee
where $\hat r(\xi,\zeta):\K\to\K$ 
is a doubly-meromorphic operator to be determined. 

\medskip

Using, the ansatz \eqref{ransatz}, we can express the 
right-hand-side of the  condition \eqref{rLax} 
as
  $$ \left\{ \tr(\L(\xi)(\si_1)x),\tr(\L(\zeta)(\si_2)y)\right\}=$$\be =-\tr\left(L(\xi)(\si_1)[x,\hat r(\xi,\zeta)y]+L(\zeta)(\si_1)[\hat r(\zeta,\xi)x,y]\right)\delta(\sigma_1-\sigma_2) -\tr\left(x\hat r(\xi,\zeta)y+y \hat r(\zeta,\xi)x \right)\delta'(\sigma_1-\sigma_2).\label{LxLybis}\ee
   To determine the  operator $\hat r(\xi,\zeta):\K\to \K$,
   it is sufficient to insert the formula \eqref{Laxfh} into Eq. \eqref{LxLybis} and to compare the result with the formula \eqref{LxLy}. This procedure works well and the result is
      \be \hat r(\xi,\zeta)=v(\xi,\zeta){\rm Id}-\frac{h(\zeta)}{4\ka}e^{\rho_r R},\label{biYBWZrmatrix}\ee
      where
    \be v(\xi,\zeta)= \frac{h(\zeta)}{4\ka}\times\frac{h(\xi)f(\zeta)+h(\zeta)f(\xi)+4\ka(\al+\al^{-1}) f(\xi)f(\zeta)+2\cos{(\rho_r)}f(\xi)}{f(\xi)-f(\zeta)}\label{v}\ee
    or, equivalently,
      \be v(\xi,\zeta)= \frac{h(\zeta)}{4\ka}\times\frac{f_-(\xi)f_-(\zeta)-f_+(\xi)f_+(\zeta)+2\cos{(\rho_r)}(f_+(\xi)+f_-(\xi))  }{f_+(\xi)+f_-(\xi)-f_+(\zeta)-f_-(\zeta)}.\label{v'}\ee
    Using the identity \eqref{id} and rewriting the identity \eqref{fid} as
              \be  h^2(\xi)+(4\ka)^2 f^2(\xi)+4\ka(\al+\al^{-1})h(\xi)f(\xi)+8\ka \cos{(\rho_l)}f(\xi)+2\cos{(\rho_r)}h(\xi)+1=0,\label{pqid}\ee
              the reader may easily compare the $\delta'(\sigma_1-\sigma_2)$, the $Y_1$ and the $Y$ terms of Eqs.\eqref{LxLy} and \eqref{LxLybis} to convince himself that the operator  $\hat r(\xi,\zeta)$ given by Eq.\eqref{biYBWZrmatrix} indeed does the job. The  comparison of the three terms gives three conditions
       \be v(\xi,\zeta)+v(\zeta,\xi)=\frac{1}{4\ka}\left((4\ka)^2 f(\xi)f(\zeta)-h(\xi)h(\zeta)-1\right).\label{delta'term}\ee
        \be f(\xi)v(\xi,\zeta)+f(\zeta)v(\zeta,\xi)=-2\cos{(\rho_l)}f(\xi)f(\zeta)-\frac{f(\xi)+f(\zeta)}{4\ka},\label{U1term}\ee
       \be h(\xi)v(\xi,\zeta)+h(\zeta)v(\zeta,\xi)=\frac{1}{2\ka} \cos{(\rho_r)}h(\xi)h(\zeta)\label{Uterm}\ee
and it is easy to check that the doubly-meromorphic function $v(\xi,\zeta)$ defined by Eq. \eqref{v'} verifies all of them. 

\medskip

The strong integrability of the bi-YB-WZ model is thus established.
   
 \section{Special limits}
The bi-YB-WZ operator $\hat r(\xi,\zeta)$ given by Eq. \eqref{biYBWZrmatrix} contains as special limits the bi-YB, the YB-WZ and the YB  $r$-matrices obtained previously in the literature. Let us give more details how this comes about.
 
 \medskip
 
 We know already that by setting \be \rho_r=2\ka b_r,\quad \rho_l=2\ka b_l,\quad \al=e^{-2\ka a},\label{para}\ee
and subsequently  taking limit $\ka\to 0$, we recover from  the bi-YB-WZ model \eqref{biYBWZbis} the bi-Yang-Baxter integrable deformation of the principal chiral model :
 \be S_{\rm bi-YB}(m)=-\int d\tau\oint \tr\biggl( m^{-1}\partial_+ m \left(a+b_rR_m +b_lR\right)^{-1}m^{-1}\partial_- m\biggr).\label{biYBbis}\ee
The $\ka\to 0$ limit of the bi-YB-WZ Lax connection \eqref{Laxconnectiontris} gives the bi-YB Lax connection found in Ref. \cite{K14}
  \be L_\pm(\xi):=\left( b_r R\mp \hat f_\pm(\xi)\right)\hat Y_\pm,
 \label{LaxconnectionbiYB}\ee
where  \be \hat Y_\pm=\left(\mp a+b_r R +b_l  R_{m^{-1}}\right)^{-1}\d_\pm mm^{-1}\label{hatYdef}\ee
and
  \be \hat f_\pm(\xi)=a\mp \ri b_l+ \left(\pm 2b_l\ri +\frac{b_l^2-b_r^2-a^2}{a}\right)\frac{1}{1\pm \xi}.\label{hatfdef}\ee
  The meromorphic functions $\hat f_\pm(\xi)$ satisfy the following  identity  
 \be \hat f_+(\xi)\hat f_-(\xi)=\hat f_+(\xi)\hat f_-(-1)+\hat f_+(+1)\hat f_-(\xi)+b_r^2.\label{hatfid}\ee
 We  find  from Eq. \eqref{biYBWZrmatrix}  that  the $\ka\to 0$ limit of the bi-YB-WZ operator $\hat r(\xi,\zeta):\K\to\K$ reads  
 \be \hat r_{\rm bi-YB}(\xi,\zeta)=\jp\left(\frac{\hat f_+(\zeta)+\hat f_-(\zeta)}{2a}-1\right)
 \left(\frac{\hat f_-(\xi)\hat f_-(\zeta)-\hat f_+(\xi)\hat f_+(\zeta)}{\hat f_+(\xi)+\hat f_-(\xi)-\hat f_+(\zeta)-\hat f_-(\zeta)}{\rm Id}-b_r R\right).\label{rbiYB}\ee
  This formula matches perfectly the last formula of Section
 6 of Ref. \cite{DLMV16} where the 
 strong integrability of the bi-YB deformation of the principal chiral model was first established. To see
 it, we must relate the parameters appearing respectively in our action \eqref{biYB} and in that of Ref. \cite{DLMV16}
 \be -2\frac{b_l}{a}=\eta,\quad -2\frac{b_r}{a}=\tilde\eta,\quad -\frac{1}{a}=K\ee
as well as relate  the spectral parameters
 \be \xi=\frac{\eta \ri +1+\js(\tilde\eta^2-\eta^2)+\nu z}{\eta \ri +1+\js(\tilde\eta^2-\eta^2)-\nu z},\  \zeta=\frac{\eta \ri +1+\js(\tilde\eta^2-\eta^2)+\nu z'}{\eta \ri +1+\js(\tilde\eta^2-\eta^2)-\nu z'},\  \nu=\sqrt{1+\jp\eta^2+\jp\tilde\eta^2 +\frac{1}{16}(\eta^2-\tilde\eta^2)^2}.\ee
 Note for completeness, that it holds
 \be \hat f_\pm (\xi(z))=\frac{\tilde\eta^2-\eta^2-4}{8K} -\frac{\nu}{2K}z^{\pm 1}.\ee
 
 \medskip
 
 If we set $b_r=0$ in the formulae \eqref{LaxconnectionbiYB} up to  \eqref{rbiYB}, we are in the setting of the Yang-Baxter deformation of the principal chiral model \eqref{YB}. We can now make comparison with
 Ref. \cite{DMV13}, where the strong integrability of this theory was first established. For that, we have to relate our notations to those  of Ref. \cite{DMV13}
 \be \frac{b_l}{a}=-\frac{\epsilon}{\sqrt{1-\epsilon^2}}, \quad a=2(1-\epsilon^2)^2,\quad \xi=\frac{\sqrt{1-\epsilon^2}+\ri\epsilon\lm}{\sqrt{1-\epsilon^2}\lm+\ri\epsilon},\quad  \zeta=\frac{\sqrt{1-\epsilon^2}+\ri\epsilon\mu}{\sqrt{1-\epsilon^2}\mu+\ri\epsilon}.\ee
 We then find
 \be \hat f_\pm(\xi(\lm))=\frac{2(1-\epsilon^2)}{1\pm\lm}\ee
 and, from Eq. \eqref{rbiYB},
 \be \hat r_{\rm YB}(\xi(\lm),\zeta(\mu))=\frac{(1-\epsilon^2)\mu^2+\epsilon^2}{1-\mu^2}\frac{1}{\lm-\mu}{\rm Id},\ee
 which indeed coincides with the result of Ref. \cite{DMV13}.
 
 \medskip
 
 It remains to consider the action of the YB-WZ model \eqref{YBWZ} which is obtained from that of the bi-YB-WZ model by setting $\rho_r=0$. In what follows, when writing the quantities $f_\pm(\xi)$, we have in mind this special case. To make comparison with
  Ref. \cite{DMV15}, where the strong integrability of the YB-WZ model was first established, we note that our parameters $\ka,\rho_l,\al$ are related to the parameters $k,K,A,\eta$ of \cite{DMV15} as
  \be \frac{\al+\al^{-1}-2\cos{\rho_l}}{\al-\al^{-1}}=k,\quad \frac{-2\ka(\al-\al^{-1})}{\al+\al^{-1}-2\cos{\rho_l}}=K, \quad \frac{-2\sin{\rho_l}}{\al-\al^{-1}}=A, \quad \frac{2\al(1-\cos{\rho_l})}{(\al-1)^2}=\eta^2\ee
and the spectral parameters are related as
  \be \xi=\frac{(f_+(0)-1)z+ 1-f_+(\infty)}{(1-f_+(\infty))z +f_+(0)-1},
  \quad  \zeta=\frac{(f_+(0)-1)z'+ 1-f_+(\infty)}{(1-f_+(\infty))z' +f_+(0)-1}.\ee
  Rewriting our formula \eqref{biYBWZrmatrix} in terms of the notation of Ref. \cite{DMV15} gives
  \be \hat r_{\rm YB-WZ}(\xi(z),\zeta(z'))=\frac{A^2+(z'-k)^2}{K(1-z'^2)}\frac{1}{z-z'}{\rm Id}.\ee
  This coincides with the result of Ref. \cite{DMV15}.
 
  \section{Conclusions and outlook}
  
  In this paper, we have proved  the strong integrability of the bi-YB-WZ $\sigma$-model. The crucial technical tool for achieving this goal
  consisted in  expressing the first order Hamiltonian dynamics of the $\sigma$-model
  in terms of a suitable non-degenerate $\E$-model. 
  
 \medskip
 
 As far as future perspectives  are concerned, we first note that the $\E$-model insight is particularly well suited for understanding   T-duality properties of the bi-YB-WZ model. In particular, we expect that the T-duality pattern of the YB-WZ model worked out in Ref. \cite{DDST18}
 could be generalized into the context of the bi-YB-WZ model. 
 
 \medskip
 
 The big challenge remains the quantization of the bi-YB-WZ model; the principal difficulty being the non-ultralocality of the Lax operator. 
 A possibility to move forward  would consist in a bi-YB-WZ generalization  of the program explored in the bi-YB context in Ref. \cite{BKL,Ko} were the monodromy matrix satisfying the  standard
Yang-Baxter Poisson algebra could be obtained working with  non-ultralocal Lax connections. 
  We expect also that other quantum insights  could  be learned  by studying the weak/strong   duality between quantum Yang-Baxter deformed $\sigma$-models and Toda-like quantum field theories as it was done e.g. in Refs. \cite{FaLi } and \cite{LS}. 
   
\end{document}